\documentclass[12pt,preprint]{aastex}

\def\hho  {H$_2$O}
\def\SgrA {Sgr~A*}
\def\SgrB  {Sgr~B2}
\def\SgrBN {Sgr~B2N}
\def\SgrBM {Sgr~B2M}

\def\c    {$\sqrt{}$}
\def\x    {\phantom{$\sqrt{}$}}
\def\p    {\phantom{0}}

\def\kms  {km~s$^{-1}$}

\def\masy {mas~y$^{-1}$}

\def\uas  {$\mu$as}
\def\deg  {\ifmmode {^\circ}\else {$^\circ$}\fi}
\def\porm {\ifmmode {\pm}\else {$\pm$}\fi}
\def\chisqpdf {\ifmmode {\chi^2_{\rm pdf}}\else {$\chi^2_{\rm pdf}$}\fi}
\def\chisq    {\ifmmode {\chi^2}\else {$\chi^2$}\fi}

\def\etal {et al.~}
\def\eg   {e.g.~}
\def\ie   {i.e.~}

\def\d    {\ifmmode {{\rlap{.}}^\circ}\else {${\rlap{.}}^\circ$}\fi}
\def\s    {\ifmmode {{\rlap{.}}^s}\else {${\rlap{.}}^s$}\fi}
\def\as   {\ifmmode {{\rlap{.}}^{''}}\else {${\rlap{.}}^{''}$}\fi}

\def\pa    {\ifmmode {\psi} \else {$\psi$}\fi}

\def\vlsr  {\ifmmode {v_{\rm LSR}}\else {$v_{\rm LSR}$}\fi}
\def\vlsrr {\ifmmode {v^r_{\rm LSR}}\else {$v^r_{\rm LSR}$}\fi}
\def\vhelio{\ifmmode {v_{Helio}}\else {$v_{Helio}$}\fi}

\def\ura   {\ifmmode {\mu_\alpha}\else {$\mu_\alpha$}\fi}
\def\udec  {\ifmmode {\mu_\delta}\else {$\mu_\delta$}\fi}

\def\uGal  {\ifmmode {\mu_{Gal}}\else {$\mu_{Gal}$}\fi}
\def\ul    {\ifmmode {\mu_l}\else {$\mu_l$}\fi}
\def\ub    {\ifmmode {\mu_b}\else {$\mu_b$}\fi}

\def\uml   {\ifmmode {v_{gr}}\else {$v_{gr}$}\fi}
\def\umb   {\ifmmode {v_b}\else {$v_b$}\fi}
\def\vsrad {\ifmmode {v_{rad}}\else {$v_{rad}$}\fi}

\def\upl   {\ifmmode {v^p_{gr}}\else {$v^p_{gr}$}\fi}
\def\upb   {\ifmmode {v^p_b}\else {$v^p_b$}\fi}
\def\vprad {\ifmmode {v^p_{rad}}\else {$v^p_{rad}$}\fi}

\def\Vo    {\ifmmode {V^{Std}_\odot}\else {$V^{Std}_\odot$}\fi}
\def\Uo    {\ifmmode {U^{Std}_\odot}\else {$U^{Std}_\odot$}\fi}
\def\Wo    {\ifmmode {W^{Std}_\odot}\else {$W^{Std}_\odot$}\fi}
\def\VH    {\ifmmode {V^H_\odot}\else {$V^H_\odot$}\fi}
\def\UH    {\ifmmode {U^H_\odot}\else {$U^H_\odot$}\fi}
\def\WH    {\ifmmode {W^H_\odot}\else {$W^H_\odot$}\fi}
\def\V     {\ifmmode {V_\odot}\else {$V_\odot$}\fi}
\def\U     {\ifmmode {U_\odot}\else {$U_\odot$}\fi}
\def\W     {\ifmmode {W_\odot}\else {$W_\odot$}\fi}

\def\Vs    {\ifmmode {V_s}\else {$V_s$}\fi}
\def\Us    {\ifmmode {U_s}\else {$U_s$}\fi}
\def\Ws    {\ifmmode {W_s}\else {$W_s$}\fi}

\def\Vsbar {\ifmmode {\overline{V_s}}\else {$\overline{V_s}$}\fi}
\def\Usbar {\ifmmode {\overline{U_s}}\else {$\overline{U_s}$}\fi}
\def\Wsbar {\ifmmode {\overline{W_s}}\else {$\overline{W_s}$}\fi}

\def\pars  {\ifmmode{\pi_s}\else{$\pi_s$}\fi}

\def\Ts    {\ifmmode{\Theta_s}\else{$\Theta_s$}\fi}
\def\Tdot  {\ifmmode{d\Theta\over dR}\else{$d\Theta\over dR$}\fi}

\def\Rp    {\ifmmode{R_p}\else{$R_p$}\fi}

\def\To    {\ifmmode{\Theta_0}\else{$\Theta_0$}\fi}
\def\Ro    {\ifmmode{R_0}\else{$R_0$}\fi}

\def\Ho    {\ifmmode{H_0}\else{$H_0$}\fi}
\def\kmspermpc  {km s$^{-1}$ Mpc$^{-1}$}

\def\Dx    {\ifmmode{\Delta x}\else{$\Delta x$}\fi}
\def\Dy    {\ifmmode{\Delta y}\else{$\Delta y$}\fi}

\shorttitle{Parallax of Sgr B2} 
\shortauthors{Reid \etal}

\begin{document}

\title{A Trigonometric Parallax of Sgr B2}

\author{M. J. Reid\altaffilmark{1}, K. M. Menten\altaffilmark{2}, 
        X. W. Zheng\altaffilmark{3}, A. Brunthaler\altaffilmark{2},  
        Y. Xu\altaffilmark{4}
       }

\altaffiltext{1}{Harvard-Smithsonian Center for
   Astrophysics, 60 Garden Street, Cambridge, MA 02138, USA}
\altaffiltext{2}{Max-Planck-Institut f\"ur Radioastronomie, 
   Auf dem H\"ugel 69, 53121 Bonn, Germany}
\altaffiltext{3}{Department of Astronomy, Nanjing University
   Nanjing 210093, China} 
\altaffiltext{4}{Purple Mountain Observatory, Chinese Academy of
   Sciences, Nanjing 210008, China}

\begin{abstract}
We have measured the positions of \hho\ masers in \SgrB, a massive star forming 
region in the Galactic center, relative to an extragalactic radio source with
the Very Long Baseline Array.  The positions measured at 12 epochs over a 
time span of one year yield the trigonometric parallax of \SgrB\ and hence a 
distance to the Galactic center of $\Ro=7.9^{+0.8}_{-0.7}$~kpc.
The proper motion of \SgrB\ relative to \SgrA\ suggests that \SgrB\ is
$\approx0.13$~kpc nearer than the Galactic center, assuming a
low-eccentricity Galactic orbit.

\end{abstract}

\keywords{Galaxy: fundamental parameters, structure, kinematics and dynamics, halo  --- 
stars: formation --- astrometry }

\section{Introduction}
        Since the distance to the Galactic Center (\Ro) was first
estimated by \citet{Shapley:18}, astronomers have expended considerable effort 
to measure \Ro\ accurately, and for good reason.  
All kinematic distances are proportional to \Ro, and hence masses and 
luminosities of giant molecular clouds and their often stars depend on directly
on \Ro.  Most luminosity and many mass estimates (\eg based on column densities) 
scale as the square of the source distance, while masses based on total 
densities or orbit fitting (using proper motions) scale as the cube of distance.  
For example, the estimate of the mass of the super-massive black hole at the 
Galactic Center (\SgrA) obtained from stellar orbits is dominated by uncertainty 
in \Ro\ \citep{Ghez:08,Gillessen:09}.

        An accurate value for \Ro\ can also have significant impact in
cosmology.  Estimates of \Ro\ and \To, the circular rotation speed of the Galaxy,
are highly correlated.  For example, measurement of the {\it apparent} 
secular motion of Sgr A* \citep{Reid:04}, caused by the orbit of the 
Sun around the Galaxy, yields the ratio \To/\Ro\ directly.  Thus, an accurate 
measurement of \Ro, will give a correspondingly accurate estimate of \To.
The value of \To\ is critical for estimating the dark matter content of 
the Milky Way, to determine if the LMC is bound to the Milky Way \citep{Shattow:08},
and, by changing the Andromeda infall speed, the total mass in the Local Group  
\citep{Oort:75,Trimble:86}. 

	The value of the Hubble constant, \Ho, is the single most 
important parameter for determining the age and size of the Universe and,
with cosmic microwave background fluctuations, the amount of dark matter and the 
equation of state of dark energy.  Currently, \Ho\ is estimated to be 
$72 \pm 3 {\rm ~(statistical)} \pm 7 {\rm ~(systematic)}$~\kmspermpc\ 
\citep{Freedman:01}.  The $\approx10$\% systematic uncertainty can be traced 
to the uncertainty in the assumed distance to the Large Magellanic Cloud.
The accurate maser distance to NGC~4258, currently accurate to $\pm7$\% 
\citep{Herrnstein:99}, coupled with HST observations of Cepheids 
in that galaxy, provides a second anchor for the extragalactic distance scale 
and reduces the systematic error in \Ho.  An accurate distance to the Galactic 
center could add a valuable third anchor for the extragalactic distance scale.
By measuring \Ro\ directly and with high accuracy, the spatial distributions of 
stars such as Cepheids could be used to re-calibrate the zero-point of their 
period-luminosity relation, the reverse of the standard procedure where their 
magnitudes are assumed and their distributions used to determine \Ro\ 
\citep{Reid:93}, and hence reinforce the foundations of the extragalactic
distance scale.

        Measurement of \Ro\ is also important to the observational study of 
strong gravitational fields.  Timing observations of the binary pulsar 1913+16 
have placed important limits on gravitational waves.  \citet{Damour:91} find 
that the largest uncertainty in modeling the orbital change of the binary 
comes from the acceleration of the Sun in its orbit about the center of the 
Galaxy (\ie \To$^2$/\Ro).   Thus, improvement in our knowledge of the 
fundamental parameters of the Milky Way are crucial to improve the accuracy 
of important tests of strong gravitational fields.

     \Ro\ has been estimated to be $8.0\pm0.5$~kpc ($\approx6$\% accuracy) 
from an ensemble of classical techniques summarized by \citet{Reid:93}.  
Since then, many more measurements of \Ro\ using these techniques have been 
published, spanning the range 7.2~\citep{Bica:06} to 8.7~kpc 
\citep{Vanhollebeke:09}.  Most classical techniques involve determining the 
distributions of large numbers of bright sources that are assumed to be 
symmetrically distributed about the Galactic center.  Distances to these sources 
are mostly photometric, which require accurate knowledge of absolute magnitudes 
and detailed calibration of the effects of metallicity, extinction and crowding, 
and combined systematic uncertainties arguably are at least 5\% and possibly
larger.  

     Significant improvement in measuring \Ro\ is likely to come from
techniques that are relatively direct.  For example, the measurement of 
stellar orbits in the Galactic center are not affected by most of the above 
mentioned systematic errors (but are sensitive to stellar crowding problems) and 
have yielded estimates of \Ro\ between 8.0 and 8.4~kpc, with uncertainties
of about 0.4~kpc \citep{Ghez:08,Gillessen:09}.  However, the most 
straight-forward and direct technique for measuring distance in astronomy 
is trigonometric parallax.  Parallaxes with accuracies better than $\pm10$~\uas\ 
have recently been obtained with Very Long Baseline Interferometric (VLBI) techniques 
\citep{Xu:06,Honma:07,Hachisuka:09,Reid:09a}, and modeling the Galaxy with
the full phase-space information provided by these observations holds great
promise to determine \Ro\ and \To\ \citep{Reid:09b}.

We are now in a position to measure the distance to the Galactic center by 
trigonometric parallax.  
In this paper we present the results of one year's observations with the
National Radio Astronomy Observatory's 
\footnote{The National Radio Astronomy Observatory is a facility of the 
National Science Foundation operated under cooperative agreement by 
Associated Universities, Inc.}
Very Long Baseline Array (VLBA) of \hho\ masers in Sagittarius~B2 (\SgrB), 
the dominant high-mass star-forming region in the Galactic center region 
\citep{Snyder:94,Belloche:08}.  These data yield the first trigonometric 
parallax for the Galactic center.  The measurement uncertainty
from only one year's data is $\pm10$\%, demonstrating the
promise of better accuracy with continued observation.

\section{Observations and Data Analysis} \label{sect:observations}

Our observations were conducted under VLBA program BR121.
We observed \SgrBM\ and \SgrBN\ over 8-hour tracks at 12 epochs
that spanned about 1 year (see Table~\ref{table:observations}
for details).  The observing dates were chosen to sample only the peaks of 
the sinusoidal trigonometric parallax signature in right ascension, 
since the declination parallax amplitude is only $\approx10$\% as great.
This sampling provides maximum sensitivity for parallax detection
and ensures that we can separate the secular proper motion 
(caused by projections of Galactic rotation, as well as any peculiar 
motion of the masers and the Sun) from the sinusoidal parallax effect.
  
\begin{deluxetable}{lllcrlc}
\tablecolumns{7} \tablewidth{0pc} \tablecaption{VLBA Observations}
\tablehead {
  \colhead{Program} & \colhead{Date} &  \colhead{Antennas Available} 
\\
  \colhead{}   & \colhead{} &  \colhead{BR FD HN KP LA MK NL OV PT SC} 
            }
\startdata
 BR121A ...    & 2006 Sep. \p4  & \c~~~\c~~~\c~~~\c~~~\c~~~\c~~~\c~~~\c~~~\c~~~\c \\
 BR121B ...    & 2006 Sep.  23  & \c~~~\c~~~\x~~~\c~~~\c~~~\c~~~\c~~~\c~~~\c~~~\c \\
 BR121C ...    & 2006 Oct. \p9  & \c~~~\c~~~\c~~~\c~~~\c~~~\c~~~\c~~~\c~~~\c~~~\c \\
 BR121D ...    & 2006 Oct.  24  & \c~~~\c~~~\c~~~\c~~~\c~~~\c~~~\c~~~\c~~~\c~~~\x \\
 BR121E ...    & 2007 Mar.  10  & \c~~~\c~~~\x~~~\c~~~\c~~~\c~~~\c~~~\c~~~\c~~~\c \\
 BR121F ...    & 2007 Mar.  17  & \c~~~\c~~~\x~~~\c~~~\c~~~\c~~~\c~~~\c~~~\c~~~\c \\
 BR121G ...    & 2007 Mar.  25  & \c~~~\c~~~\c~~~\c~~~\c~~~\c~~~\c~~~\x~~~\c~~~\c \\
 BR121H ...    & 2007 Apr. \p4  & \c~~~\c~~~\c~~~\c~~~\c~~~\c~~~\c~~~\c~~~\c~~~\c \\
 BR121I .....  & 2007 Apr.  22  & \c~~~\c~~~\c~~~\c~~~\c~~~\c~~~\c~~~\c~~~\c~~~\c \\
 BR121M ...    & 2007 Sep.  28  & \c~~~\x~~~\c~~~\c~~~\c~~~\c~~~\c~~~\c~~~\c~~~\x \\
 BR121K ...    & 2007 Sep.  30  & \c~~~\c~~~\c~~~\c~~~\c~~~\c~~~\c~~~\c~~~\c~~~\x \\
 BR121N ...    & 2007 Oct.  16  & \c~~~\c~~~\c~~~\c~~~\c~~~\c~~~\c~~~\c~~~\c~~~\x \\
\enddata
\tablecomments {Check marks indicate that that antenna produced good data, while
a blank indicates that little or no useful data was obtained.  Antenna codes are
BR: Brewster, WA; FD: Fort Davis, TX; HN: Hancock, NH; KP: Kitt Peak, AZ;
LA: Los Alamos, NM; MK: Mauna Kea, HI; NL: North Liberty, IA; OV: Owens Valley, CA;
PT: Pie Town, NM; and SC: Saint Croix, VI.
Note that epoch
M preceded epoch K as indicated.  Ultimately,
owing to scatter broadening of the image of J1745--2820, interferometer baselines 
longer than 2000~km could not be effectively used, limiting the usefulness of
the HN, MK and SC antennas for this study.
  }
\label{table:observations} 
\end{deluxetable}

\begin{figure}[htp]
\includegraphics{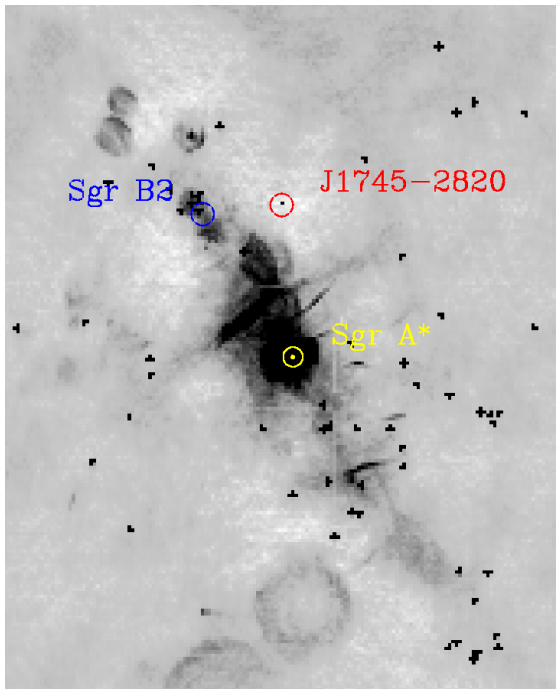}
\caption{\small
  VLA 90-cm wavelength image, adapted from \citet{LaRosa:00}, 
  with the locations of \SgrA, \SgrB, and J1745--2820 indicated. 
  The separation between \SgrB\ and J1745--2820 is $\approx20'$. 
        }
\label{fig:overlay}
\end{figure}

For both \SgrBM\ and \SgrBN\ we used the background source, 
J1745--2820, as a position reference for the parallax and proper motion 
measurements (see Figure~\ref{fig:overlay}).  
J1745--2820 is a well-studied extragalactic radio
source \citep{Bower:01} that has been used in previous 
VLBA astrometric observations of \SgrA\ \citep{Reid:04}.
Table~\ref{table:positions} lists the positions of these sources. 
J1745--2820 is projected only $20'$ west of the maser
sources, making it a nearly ideal position reference as it is very close
to our maser targets, thereby canceling most systematic errors by a factor
of 0.006 (the angular separation in radians).  Also, as its offset is predominantly
in the east-west direction, it samples similar source zenith angles as the target
masers, which further reduces the effects of unmodeled atmospheric delays
\citep{Honma:08}.  

We alternated between two 16~min observing blocks, one for \SgrBM\ and
the other for \SgrBN; each block consisted of observations of a maser target 
and J1745--2820.  Within a block, we switched sources every 40~s during the first
two epochs and every 15~s (in order to better monitor and remove rapid atmospheric 
phase fluctuations) during the remaining 10 epochs. 
We used a strong \hho\ maser spot as the interferometer phase-reference, 
because it was considerably stronger than the background source and could be 
detected on individual baselines in the available on-source time as short
as 8~s.  

We placed observations of strong sources near the beginning, 
middle and end of the observations in order to 
monitor delay and electronic phase differences among the IF bands.
In practice, however, we found minimal drifts and used only a single
scan on J1642+4938 for this calibration.  We did not attempt
bandpass calibration as the variation in phase across the VLBA bandpasses
are typically  $<5^\circ$ across the central 90\% of the band, and
the masers were observed near band center.  We also placed $\approx40$~min
``geodetic-like'' observing blocks before, near the middle, and after the
rapid switching blocks in order to monitor and remove the effects of
zenith path-length errors in the VLBA correlator model \citep{Reid:09a}.

The rapid-switching observations employed four adjacent frequency bands of
8 MHz bandwidth and recorded both right and left circularly polarized
signals.  The four (dual-polarized) bands were centered at 
Local Standard of Rest velocities ($\vlsr$) of
158, 50, $-58$ and $-166$ \kms\ for both maser sources, with almost all
known \SgrB\ maser signals contained in the second band \citep{McGrath:04}.
The raw data were recorded on transportable disks
at each antenna and shipped to the VLBA correlation facility in
Socorro, NM.  The data from individual antenna pairs were cross-correlated 
with an integration time of 0.52~s.  Integration times were 
kept short to allow position shifting of the data to accommodate a priori
uncertainties in the maser positions.  With this short integration time,
we could not process all eight frequency bands in one pass with 
sufficient spectral resolution for the masers, without 
exceeding the maximum correlator output rate.  Thus, we correlated 
the data in two passes.  One pass was processed with 16 spectral 
channels for each of the eight frequency bands.  This data was used for 
the geodetic blocks (to determine atmospheric delays and clock 
drifts) and for the background continuum sources observed in 
rapid-switching (phase-referencing) mode.  Another pass was 
processed with 256 spectral channels, but only for the single 
(dual-polarized) frequency band containing the maser signals,
giving spectral channels separated by 0.42~\kms, assuming a rest frequency of 
22235.08~MHz for the $6_{16} \rightarrow 5_{23}$ transition of \hho.

\begin{deluxetable}{lllcrlc}
\tablecolumns{7} \tablewidth{0pc} \tablecaption{Source Characteristics}
\tablehead {
  \colhead{Source} & \colhead{R.A. (J2000)} &  \colhead{Dec. (J2000)} &
  \colhead{$\theta_{\rm sep}$} & \colhead{P.A.} &
  \colhead{$S_\nu$} & \colhead{\vlsr} 
\\
  \colhead{}       & \colhead{(h~~m~~s)} &  \colhead{(d~~'~~'')} &
  \colhead{(deg)} & \colhead{(deg)} &
  \colhead{(Jy)} & \colhead{(\kms)}
            }
\startdata
 \SgrBM\ ...    & 17 47 20.150  &--28 23 04.03  & ...  & ...   &40   & 66.4 \\
 J1745--2820 ... & 17 45 52.4968 &--28 20 26.294 & 0.32 & $-82$ &0.1  &... \\
 \\
 \SgrBN\ ...    & 17 47 19.926  &--28 22 19.37  & ...  & ...   &13   & 56.7 \\
 J1745--2820 ... & 17 45 52.4968 &--28 20 26.294 & 0.32 & $-84$ &0.1  &... \\
\enddata
\tablecomments {  
  Coordinates are those used in the VLBA correlator.  The position of J1745--2820
  is accurate to about $\pm10$~mas \citep{Reid:04}.  Maser spots in both source 
  are spread over a region of several arcseconds.  
  Angular offsets ($\theta_{\rm sep}$)  
  and position angles (P.A.) east of north of the background source (J1745--2820) 
  relative to the maser source are indicated in columns 4 and 5. 
  Typical flux densities ($S_\nu$) and LSR velocities of the maser spots used for the
  parallax measurement and of the background source are given in columns 6 and 7.
  To avoid potential confusion, we note that the listed \SgrBM\ maser spot was the 
  interferometer phase reference, whereas the listed \SgrBN\ maser spot was not
  the phase reference.  
  }
\label{table:positions} 
\end{deluxetable}

Calibration of the interferometer data was done with the Astronomical
Image Processing System (AIPS) in a manner described by \citet{Reid:09a}
for methanol maser parallaxes.  A single spectral-channel of the maser
data was used for the interferometer phase reference.  
For \SgrBM, a maser spot at $\vlsr=66.4$~\kms\ served as the phase reference and
was compact ($\approx0.3$~mas FWHM) and detectable at all epochs with typical
strength of 40~Jy.  Test imaging of this spot showed a nearly unresolved
source with high dynamic range ($>100:1$), indicating a clean reference source.  
When calculating reference phases
it is important to have very low residual fringe rates in order to avoid
degrading the images of the target source \citep{Beasley:95}, in our case 
J1745--2820.  Since the
position of J1745--2820 was known to better than $10$~mas accuracy
\citep{Reid:04}, we could measure its {\it apparent} map offset when phase
referenced to a maser spot and correct the position of the maser 
(to better than $10$~mas accuracy) before final processing.  
For the \SgrBM\ reference spot this required a position shift of
($\Delta\alpha\cos{\delta},~\Delta\delta$)=(1\as594,~--1\as399); 
the same shift was used for all
12 epochs, canceling to second-order the effects associated with
the error in this shift. (This shift should be added to the correlation
coordinates in Table~\ref{table:positions} to give the true position of the 
reference maser spot.)  

For \SgrBN, no strong maser spot suitable as a phase reference was found to 
persist for all 12 epochs.   Therefore, we used the maser spot  
at $\vlsr=52.5$~\kms\ as the reference for epochs A through I, over
which time it displayed $S_{\nu}>20$~Jy.  For epochs K through N the spot at 
$\vlsr=38.6$~\kms\ was used as the reference over which time it displayed 
$S_{\nu}>180$~Jy.  As described above, we determined position shifts
relative to the coordinates used for correlation (Table~\ref{table:positions}) 
for these two reference spots, based on trial calibration and imaging of 
J1745--2820.  The spot at $\vlsr=52.5$~\kms\ was found at 
($\Delta\alpha\cos{\delta},~\Delta\delta$)=(0\as023,~0\as028) and the spot 
at $\vlsr=38.6$~\kms\ at 
($\Delta\alpha\cos{\delta},~\Delta\delta$)=(--0\as011,~0\as006).  
The effects of these shifts were removed from the maser data for epochs A 
through I and from epochs K through N, respectively, before determining 
reference phases.

Reference phases were interpolated from those measured at the times of
the maser observations to the times of the J1745--2820 observations and 
subtracted from all data.  Phase reference
data for each baseline were examined for jumps greater than 60\deg\ between
adjacent maser scans and, when found, the J1745--2820 data between these
scans were discarded.
Calibrated data for J1745--2820 were imaged with the AIPS task IMAGR.
All data with source elevation less than 20\deg\ were discarded, owing to
greatly increased sensitivity to atmospheric delay errors compared to
higher elevation data. 
While interferometer fringes were obtained on the longest baselines for
the maser data, indicating maser spot sizes of $\approx0.3$~mas, the background
continuum source was considerably larger, probably owing to scatter broadening
\citep{Bower:01}, and we could only effectively image this source with 
baselines shorter than 1500~km.  Unfortunately, this reduced our
astrometric accuracy by a factor of $\approx4$ compared to the \SgrB\
masers which could be detected on the longest VLBA baselines.
For J1745--2820 we used a ($u,v$)-cellsize of 0.1~mas and a CLEAN Gaussian restoring 
beam with a FWHM of 1.5~mas.  Images were fitted with a single elliptical Gaussian
model using the task JMFIT. 

\section{Results} \label{sect:results}

Parallax data was obtained by subtracting the apparent position of J1745--2820
from those of a maser spot that could be detected at all epochs.  For
\SgrBM\ we used the phase-reference spot for the parallax measurement.
However, for \SgrBN, neither phase-reference spot persisted for the
entire year and we selected different maser spots for the parallax measurement.
When analyzing this data, we corrected for the
difference in the shifts used for the two different reference maser spots.
The measured position differences are given in Table~\ref{table:data}.  

Formal position fitting uncertainties are quite small 
(generally $<0.02$~mas toward the east and $<0.04$~mas toward the north ) 
for both the strong maser spots and J1745--2820 observed with the VLBA.  
However, realistic uncertainties need to account for systematic effects, 
usually dominated by residual errors in (mostly tropospheric) 
propagation delays after calibration, and possibly structural changes in the
maser spots and/or in the background source.  
Also, the typically phase-stable weather for early morning observations
in late winter/early spring produced significantly better phase
reference data than the late evening observations in late summer/early
fall.  We estimated realistic
uncertainties from the quality of the phase reference data as well as
the scatter among positions measured at closely spaced epochs.  
These uncertainties are listed in Table~\ref{table:data} and were used to
weight the data when fitting for parallax and proper motion.

\begin{figure}[htp]
\epsscale{1.0} 
\includegraphics{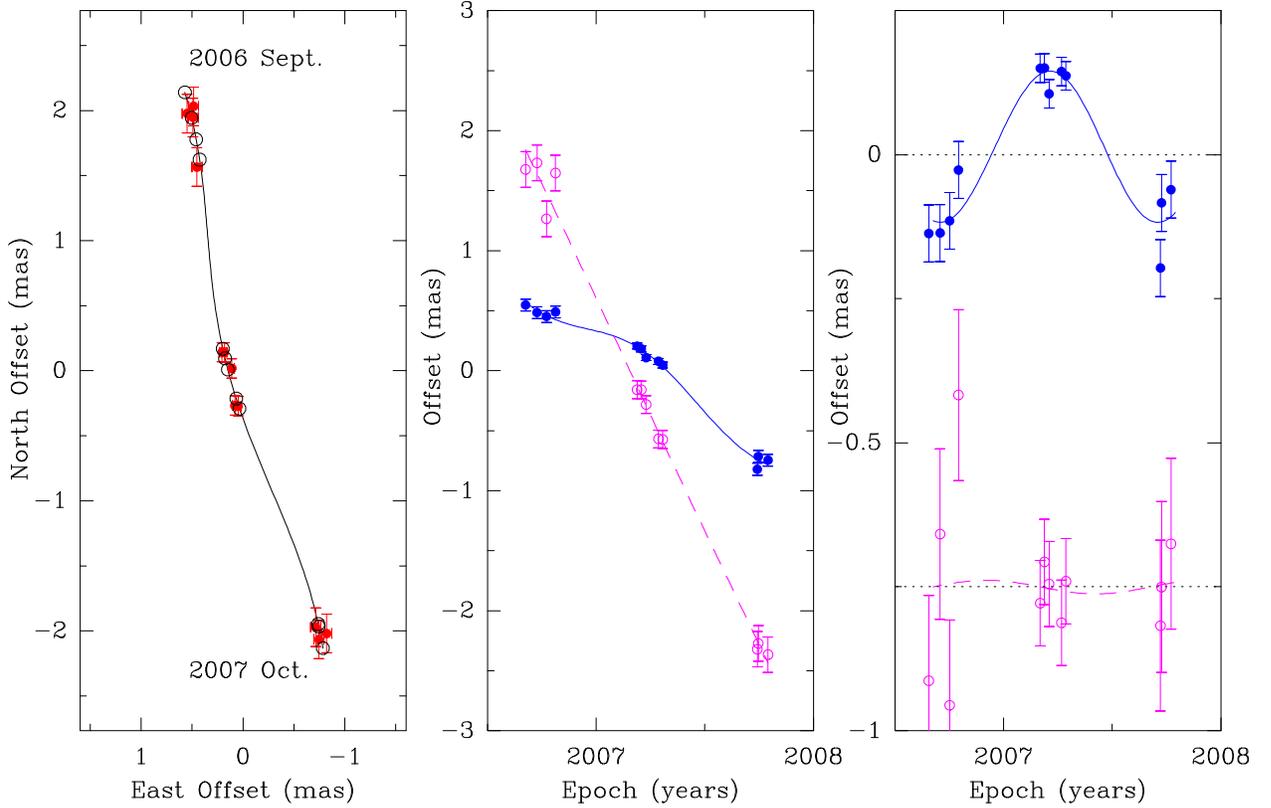}
\caption{\small
  Parallax and proper motion data and fits for \SgrBM.
  Plotted are position measurements of an \hho\ maser spot 
  at $\vlsr=66.4$~\kms\ relative to the background source J1745--2820. 
  {\it Left Panel:} Positions on the sky ({\it red circles}) 
  with first and last epochs labeled.
  The expected positions from the parallax and proper motion fit
  are indicated {\it (black circles and solid line)}.
  {\it Middle Panel:} East {\it (filled blue circles and solid line)} and 
  North {\it (open magenta circles and dashed line)}
  position offsets and best fit parallax and proper motions fit versus time.
  The northward data have been offset from the eastward data for clarity.  
  {\it Right Panel:} Same as the {\it middle panel}, except the
  best fit proper motion has been removed, allowing
  the effects of only the parallax to be seen.
        }
\label{fig:b2m_fit}
\end{figure}

\begin{figure}[htp]
\epsscale{1.0} 
\includegraphics{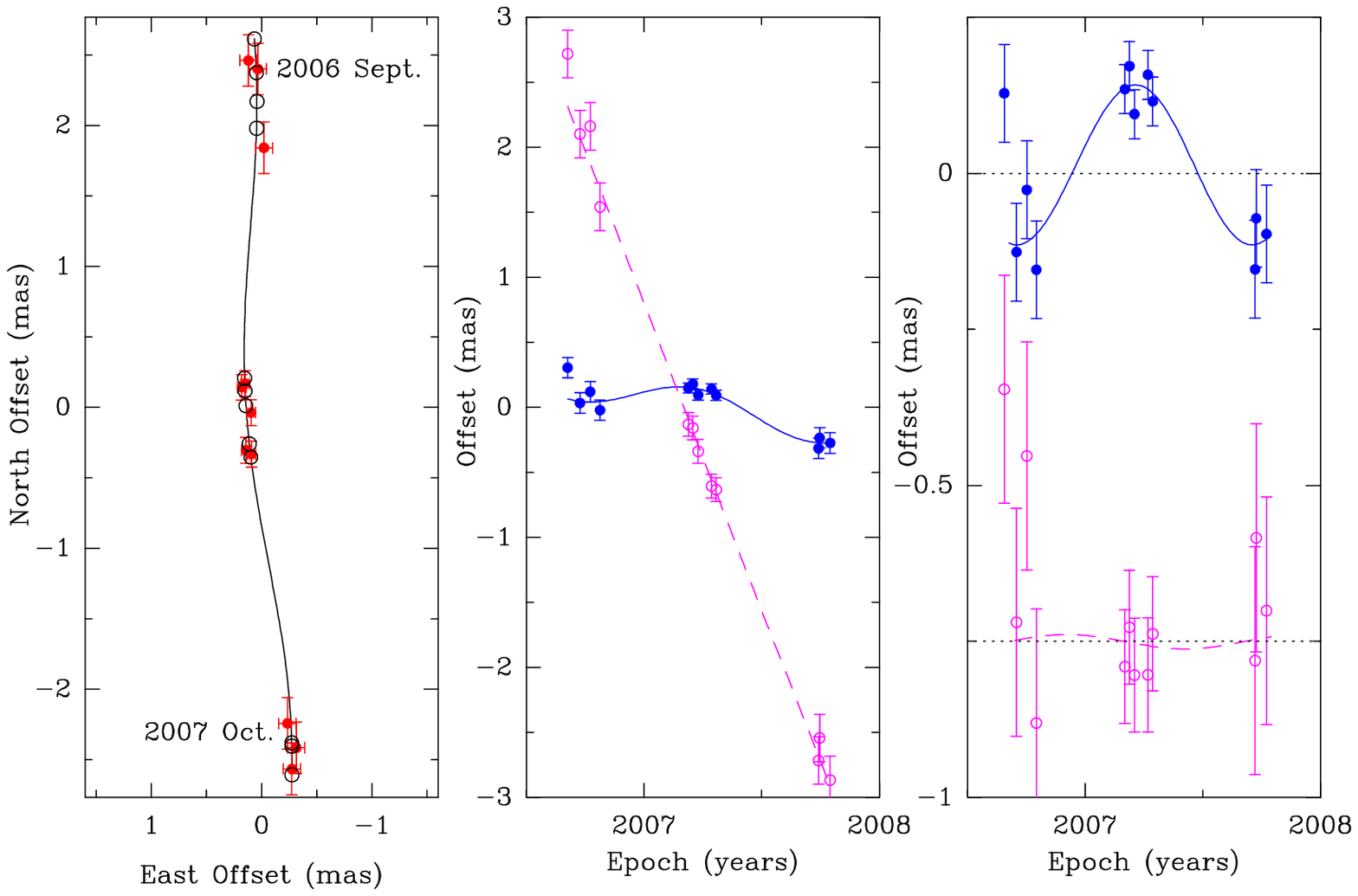}
\caption{\small
  Parallax and proper motion data and fits for \SgrBN.
  Plotted are position measurements of an \hho\ maser spot 
  at $\vlsr=56.7$~\kms\ relative to the background source J1745--2820. 
  {\it Left Panel:} Positions on the sky ({\it red circles}) 
  with first and last epochs labeled.
  The expected positions from the parallax and proper motion fit
  are indicated {\it (black circles and solid line)}.  The first 
  epoch point appears to be an outlier and was not used for the fitting.
  {\it Middle Panel:} East {\it (filled blue circles and solid line)} and 
  North {\it (open magenta circles and dashed line)}
  position offsets and best fit parallax and proper motions fit versus time.
  The northward data have been offset from the eastward data for clarity.  
  {\it Right Panel:} Same as the {\it middle panel}, except the
  best fit proper motion has been removed, allowing
  the effects of only the parallax to be seen.
        }
\label{fig:b2n_fit}
\end{figure}

The model for the data consisted of the sum of the sinusoidal parallax
signature (corrected for the eccentricity of the Earth's orbit about the
Sun) and a linear proper motion in each coordinate.  We performed 
least-squares fits separately for the \SgrBM\ and \SgrBN\ data.
Table~\ref{table:parallax_fits} and Figures~\ref{fig:b2m_fit} and \ref{fig:b2n_fit}
show the parallax and proper motion fitting results for the \SgrBM\ 
and \SgrBN\ \hho\ masers.

\begin{deluxetable}{lcrrc}
\tablecolumns{5} \tablewidth{0pc}
\tablecaption{VLBA Position Measurements}
\tablehead {
  \colhead{Source} &\colhead{\vlsr}& \colhead{Epoch} & \colhead{\Dx}   & \colhead{\Dy} 
\\
  \colhead{}       &\colhead{(\kms)}& \colhead{(yr)}  & \colhead{(mas)} & \colhead{(mas)}
           }
\footnotesize
\startdata
\SgrBM\ ...   &66.4&  2006.676 & $  +0.318\pm0.050$  & $  +2.076\pm0.150$ \\
         &&  2006.728 & $  +0.255\pm0.050$  & $  +2.131\pm0.150$ \\
         &&  2006.772 & $  +0.222\pm0.050$  & $  +1.665\pm0.150$ \\
         &&  2006.813 & $  +0.260\pm0.050$  & $  +2.046\pm0.150$ \\
         &&  2007.189 & $  -0.025\pm0.025$  & $  +0.241\pm0.075$ \\
         &&  2007.208 & $  -0.048\pm0.025$  & $  +0.240\pm0.075$ \\
         &&  2007.230 & $  -0.120\pm0.025$  & $  +0.117\pm0.075$ \\
         &&  2007.287 & $  -0.151\pm0.025$  & $  -0.169\pm0.075$ \\
         &&  2007.307 & $  -0.183\pm0.025$  & $  -0.174\pm0.075$ \\
         &&  2007.742 & $  -1.051\pm0.050$  & $  -1.921\pm0.150$ \\
         &&  2007.747 & $  -0.944\pm0.050$  & $  -1.873\pm0.150$ \\
         &&  2007.791 & $  -0.975\pm0.050$  & $  -1.967\pm0.150$ \\
\\
\SgrBN\ ...   &56.7&  2006.676 & $ 193.960\pm0.060$  & $ -33.656\pm0.140$ \\
         &&  2006.728 & $ 193.689\pm0.060$  & $ -34.273\pm0.140$ \\
         &&  2006.772 & $ 193.775\pm0.060$  & $ -34.213\pm0.140$ \\
         &&  2006.813 & $ 193.634\pm0.060$  & $ -34.833\pm0.140$ \\
         &&  2007.189 & $ 193.804\pm0.030$  & $ -36.506\pm0.070$ \\
         &&  2007.208 & $ 193.835\pm0.030$  & $ -36.532\pm0.070$ \\
         &&  2007.230 & $ 193.751\pm0.030$  & $ -36.712\pm0.070$ \\
         &&  2007.287 & $ 193.796\pm0.030$  & $ -36.979\pm0.070$ \\
         &&  2007.307 & $ 193.747\pm0.030$  & $ -37.007\pm0.070$ \\
         &&  2007.742 & $ 193.340\pm0.060$  & $ -39.090\pm0.140$ \\
         &&  2007.747 & $ 193.420\pm0.060$  & $ -38.917\pm0.140$ \\
         &&  2007.791 & $ 193.381\pm0.060$  & $ -39.240\pm0.140$ \\
\enddata
\tablecomments {
 {\footnotesize
   Data used for parallax and proper motion fits for an \hho\ maser
   spot in \SgrBM\ and \SgrBN.
   Columns 4 and 5 are position differences between the maser spot and
   a compact extragalactic source J1745--2820, relative to the
   coordinates used to correlate the VLBA data and any shifts applied
   in calibration (see Section~\ref{sect:observations}); position 
   differences are in the eastward ($\Dx=\Delta\alpha \cos{\delta}$) 
   and northward directions ($\Dy=\Delta\delta$).   The large offsets
   for the \SgrBN\ maser spot reflect its position relative to the adopted 
   phase reference at $\vlsr=52.5$~\kms.
  }
}
\label{table:data}
\end{deluxetable}

The \SgrBM\ data for the maser spot at $\vlsr=66.4$~\kms\ could be well 
fitted by the model with a parallax of $0.130\pm0.012$~mas and a post-fit 
chi-squared per degree of freedom (\chisqpdf) near unity.  At two epochs 
one of the important ``inner-5'' VLBA antennas was not available 
(OV at 2007 Mar. 25 = 2007.230 and FD at 2007 Sep. 28 = 2007.742) 
and the post-fit residuals were somewhat larger than most others.  
Removing these points resulted in a parallax of $0.127\pm0.012$~mas, 
which is very close to the parallax using all data; we adopted the
fit using all data.  We also tested the sensitivity of the parallax
to the errors assigned to the data by assuming uniform values for 
each coordinate.  We achieved a \chisqpdf\ of unity for errors of
0.037~mas and 0.143~mas for the eastward and northward data and a 
parallax of $0.132\pm0.013$~mas.  Thus, details of weighting the
data had little effect on the parallax measurement.

The \SgrBN\ data for the maser spot at $\vlsr=56.7$~\kms\ displayed a 
bit more scatter than the \SgrBM\ data.
Using all data, the parallax for \SgrBN\ was $0.111\pm0.019$~mas.
The first epoch point appeared discrepant and after removing it the
fit improved considerably, yielding a parallax of $0.128\pm0.015$~mas.
We adopt this result for \SgrBN.  Some other maser spots (spectral
channels at $\vlsr=47.9, 47.5, 47.1, 42.0~{\rm and}~41.6$~\kms)
could be detected at all epochs and yielded nearly identical parallaxes.
However, comparison of the post-fit residuals indicated that the
data from different spectral channels were highly correlated, and thus
combining the results from these channels produces no significant
improvement for the parallax result.   This does, however, indicate
that potential variations of the maser spot structure is not an important
source of systematic uncertainty.

\begin{deluxetable}{lrrrrrr}
\tablecolumns{7} \tablewidth{0pc}
\tablecaption{Parallaxes \& Proper Motions for \SgrB}
\tablehead {
  \colhead{Source} & \colhead{$\ell$} & \colhead{$b$} &
  \colhead{Parallax} & \colhead{$\mu_x$} & \colhead{$\mu_y$} &
  \colhead{\vlsr} 
\\
  \colhead{}      & \colhead{(deg)} & \colhead{(deg)} &
  \colhead{(mas)} & \colhead{(\masy)} & \colhead{(\masy)} &
  \colhead{(\kms)}         
           }
\startdata 
\SgrBN\ ...& 0.677 &$-0.028$ &$0.128\pm0.015$ &$-0.32\pm0.05$ &$-4.69\pm0.11$ &$64\pm5$ \\
\SgrBM\ ...& 0.667 &$-0.035$ &$0.130\pm0.012$ &$-1.23\pm0.04$ &$-3.84\pm0.11$ &$61\pm5$ \\
\enddata
\tablecomments {
 {\footnotesize
   Parallax and proper motion fits using the data in Table~\ref{table:data}.
   Columns 2 and 3 give Galactic longitude and latitude, respectively.  
   Columns 4, 5 and 6 list the parallax and proper motions in the eastward 
   ($\mu_x=\ura \cos{\delta}$) 
   and northward directions ($\mu_y=\udec$), respectively.  
   Column 7 lists the central Local Standard of Rest velocity of dense molecular
   gas associated with the maser sources \citep{Reid:88,Sutton:91}.
  }
}
\label{table:parallax_fits}
\end{deluxetable}

Since the \SgrBM\ and \SgrBN\ data were taken at different times
(interleaved observations; see Section~\ref{sect:observations}) and are 
{\it statistically} independent, one could consider combining the solutions and 
obtaining some improvement in parallax accuracy.  However a detailed 
comparison of the data suggests that systematic errors common to both sources 
are significant.  One can see, in the right-hand panels of 
Figures~\ref{fig:b2m_fit} and \ref{fig:b2n_fit}, similar deviations in 
post-fit residuals (departures from the fitted line) for the two maser 
sources for the five epochs in early 2007 and the 3 epochs in late 2007.   
But, the residuals for the four epochs in late 2006 show no strong 
correlation, suggesting that the results for the two maser sources are not 
entirely correlated.  However, we will be conservative when quoting
a single parallax for the \SgrB\ region and not reduce the uncertainty
when averaging the parallaxes from the two sources.  Thus, our combined 
result is that \SgrB\ region has a parallax of $0.129\pm0.012$~mas, 
corresponding to a distance of $7.8^{+0.8}_{-0.7}$~kpc.

\section{Discussion} \label{sect:discussion}

Our measurement of the distance to \SgrB\ almost certainly can be
taken as a measure of \Ro, the Galactic center distance.  
Evidence that \SgrB\ is very close to the the Galactic center
(\ie \SgrA) has been summarized by \cite{Reid:88}, and includes the
following observations:
1) \SgrB\ is a nearly unique source with the greatest variety of detected
molecular species of any molecular cloud, owing to the high densities near 
the Galactic center \citep{Snyder:94,Belloche:08}.
2) It is projected $\approx0.09$~kpc from the Galactic center,
which suggests an expected line-of-sight component of 0.07~kpc.
3) Atomic and molecular absorption studies locate \SgrB\ within
the ``270-pc expanding shell'' \citep{Scoville:72}.  

Since we have also measured the proper motions of maser spots in
\SgrB, we can estimate its 3-dimensional location, provided 
it is on a nearly circular Galactic orbit.  For an object 
near the Galactic center on low eccentricity orbit in the plane,  
the the distance offset along the line-of-sight from the center, $d$, 
is given by $d = r_{proj} \Ro \uGal / \vlsr$, where $r_{proj}$ is the projected 
distance on the sky and $\uGal$ is the proper motion in Galactic longitude
in a reference frame not rotating with the Galaxy.  
Our measured proper motions for a small number of maser 
spots in \SgrBM\ and \SgrBN\ gives an average motion in Galactic longitude of 
$-4.05$~\masy, with a probable uncertainty of about 1~\masy\ because 
we have not accounted for internal motions in the two \hho\ maser sources.
(A complete mapping of the $\sim100$ maser spots in each source for multiple
epochs is beyond the scope of this paper and will be published later.)  
Subtracting the apparent motion of \SgrA\ of --6.379~\masy\ in Galactic longitude
\citep{Reid:04} (in order to correct for the apparent motion expected from the 
solar orbit) yields a true motion in the direction of increasing Galactic 
longitude of $\uGal\approx2.3\pm1.0$~\masy.
Adopting $\Ro\approx8$~kpc and $\vlsr\approx62$~\kms, 
this suggests $d\approx0.13\pm0.06$~kpc. 
This estimate of the offset of \SgrB\ from \SgrA\ rests on the
assumption of a low eccentricity Galactic orbit for \SgrB.  This
assumption might be testable in the future when the full space motion 
of \SgrB\ is accurately measured and the gravitational potential in the 
Galactic center region is better known.
However, since the offset of \SgrB\ from \SgrA\ is almost certainly
much smaller than our (current) distance measurement uncertainty,
the exact value of the offset is probably of little significance.

Correcting for the small estimated offset of \SgrB\ relative to \SgrA\ of 
0.13~kpc (with \SgrB\ closer than \SgrA), we find $\Ro=7.9^{+0.8}_{-0.7}$~kpc.
While this estimate of \Ro\ is not (yet) as accurate as, for example, 
that obtained from stellar orbits in the Galactic center, a trigonometric 
parallax is the ``gold standard'' for distances and provides improved
confidence in any estimate of \Ro.  The uncertainty in
\Ro\ can be reduced by continued observations (with $\sigma_\Ro=1/\sqrt{N}$
for $N$ similar yearly observations).  Improved accuracy could also come
from the discovery of another suitable extragalactic reference source that is
less scatter broadened than J1745--2820.

\vskip 0.2truein 
XWZ and YX were supported by the Chinese National Science Foundation, 
through grants NSF 10673024, 10733030, 10703010 and 10621303, and by the NBPRC 
(973 Program) under grant 2007CB815403.

\end{document}